\documentclass[prb,floatfix,amsfonts,amssymb,twocolumn]{revtex4}
\usepackage{graphics}
\usepackage{dcolumn}
\usepackage{bm}
\usepackage{subfigure}
\usepackage{epsfig}

\usepackage{amsmath}
\usepackage{amsfonts}
\usepackage{amssymb}

\usepackage[colorlinks=true,linkcolor=blue]{hyperref}%
\expandafter\ifx\csname package@font\endcsname\relax\else
 \expandafter\expandafter
 \expandafter\usepackage
 \expandafter\expandafter
 \expandafter{\csname package@font\endcsname}%
\fi
\begin{document}

\title{Current-induced nucleation and motion of zero field skyrmion}


\author{Sougata Mallick}\email{sougata.physics@gmail.com}
\author{Sujit Panigrahy}
\author{Gajanan Pradhan}
\author{Stanislas Rohart}\email{stanislas.rohart@universite-paris-saclay.fr}
\affiliation{Laboratoire de Physique des Solides, Universit\'e Paris-Saclay, CNRS UMR 8502, F-91405 Orsay Cedex, France}

\date{\today}

\begin{abstract}
We study the stabilization and electrical manipulation of skyrmions in magnetic ultrathin films in the absence of an applied magnetic field. We show that this requires an increased magnetic anisotropy, controlled by the sample thickness, as compared to usual skyrmionic samples, so that the uniform state corresponds to the zero field ground state and the skyrmions to metastable excitations. Although skyrmion stabilization at zero field is demonstrated over a broad range of thicknesses, electrical control appeared to be more demanding to avoid skyrmion deformation. In the thinnest samples, the large magnetic anisotropy prevents skyrmion deformations and we show that they can be nucleated progressively by current pulses, which underlines that the only possible transition occurs between uniform and skyrmion states. The solitonic skyrmions in zero applied magnetic field have the same properties as compared to field stabilized ones, with a long-term stability and high mobility when excited by a spin-orbit torque.
\end{abstract}

\maketitle

\section{Introduction}

Magnetic skyrmions are appealing magnetic textures to develop innovating spintronic devices and applications \cite{Fert_Nat.Nano., Zhang_Sci.Rep., Song_Nat.Elec.}. In that perspective, the information process is based on the controlled nucleation and motion of individual skyrmions, that are considered as solitons\cite{Sampaio_Nat.Nano., Woo_Nat.Mat.}. Therefore, the magnetic materials should not host arrays of skyrmions, but metastable excitations over a homogeneous magnetization ground state \cite{Buttner_Sci.Rep.}. To this aim, in most studies\cite{Romming_PRL,Luchaire_Nat.Nano.,Boulle_Nat.Nano.,Woo_Nat.Mat.,Jiang_Science, Yu_Nature,Hrabec_Nat.Comm.}, the application of a magnetic field is required in order to promote a uniform ground state over which metastable skyrmions can be manipulated individually . However, as illustrated by several theoretical works, solitonic skyrmions can exist even in zero applied field, but require a fine tuning of the micromagnetic parameters\cite{Sampaio_Nat.Nano.,Rohart_PRB,Back_JPD}. From the application point of view, avoiding an applied magnetic field sound appealing to reduce the complexity and enhance the device efficiency. In the perspectives of using antiferromagnetic materials (real antiferromagnetsca\cite{barker2016}, as well as ferrimagnets\cite{Hirata_Nat.Nano.,caretta_Nat.Nano} or synthetic antiferromagnets\cite{zhang2016,Legrand_Nat.Mat.}), where samples are not sensitive to an external field, the ability to stabilize skyrmions at remanence is crucial.

The possibility of stabilizing skyrmions in the absence of an external field has received great attention, and led to different successful routes.
An elegant possibility is the substitution of the external magnetic field by a build-in internal field, induced for example by exchange bias at the interface with an antiferromagnetic layer\cite{Rana_PRA,Guang-Nat.Comm.} or by an RKKY induced coupling from a homogeneous ferromagnetic layer through a well-chosen non-magnetic spacer\cite{Legrand_Nat.Mat.}. Some samples also enable to get rid of any bias field, through either geometrical confinement\cite{Boulle_Nat.Nano.,Gallagher-PRL,Ho-PRAppl}, particular magnetic field history\cite{Duong_APL,Li_APL}, micromagnetic parameter optimization\cite{Brandao-Appl.Surf.Sci.,Bhattacharya_Nat.Elec.,caretta_Nat.Nano}, via thermal effects \cite{Lemesh-Adv.Mat.}, or strong frustration of the exchange interaction\cite{Meyer-Nat.Comm}. However, a demonstration of the solitonic character of the observed skyrmions, is yet to be demonstrated. Indeed, beyond the observation of non-trivial topological and compact domains, the control and nucleation by current of non-deformable textures would provide a real proof of the skyrmionic character, so that zero field skyrmions opens significant perspectives to the field of skyrmion-based devices.

In this article, we study the dynamics of zero field magnetic skyrmions in a Pt/Co/Au based multilayer. Close to the magnetic anisotropy compensation, samples display a spontaneous stripe demagnetization pattern at zero field, as observed in most skyrmion host materials\cite{Romming_PRL,Luchaire_Nat.Nano.,Boulle_Nat.Nano.,Woo_Nat.Mat.,Jiang_Science, Yu_Nature,Hrabec_Nat.Comm.}. When reducing the Co layer thickness, the increase of the magnetic anisotropy improves the uniform state stability and there is a thickness range when uniform, stripe and  skyrmion-like states are stable at zero field, depending on the sample magnetic history. In connected patterned samples, we study the dynamics of these excitations. The observation of zero field skyrmion-like textures, obtained from a field process is found to be a non-sufficient condition to enable their current induced control. Particularly, the ability to nucleate skyrmions directly from the uniformly magnetized state is indeed a challenging process, since no specific final state is promoted, contrary to the experiments at finite magnetic field~\cite{Fallon_Small,Jiang_Science,Hrabec_Nat.Comm.}. The variety of possible excited states implies that a high optimization of the sample parameters is required so that the skyrmion is the only outcome of a current induced nucleation experiment. In the best samples, we show that the textures are true zero field skyrmions: they can be nucleated from the uniform magnetic state using standard current nucleation techniques\cite{Fallon_Small,Jiang_Science,Hrabec_Nat.Comm.}, they obey a similar behavior as field stabilized skyrmions\cite{Woo_Nat.Mat.,Litzius_Nat.Phys.,Hrabec_Nat.Comm.,Jiang_Nat.Phys.}, including the gyrotropic deflection and velocities larger than 70~m/s.

\section{Experimental details}

Our samples are based on a Co magnetic bilayer with a symmetric stacking, following the approach of Hrabec et al.\cite{Hrabec_Nat.Comm.,Hrabec_PRL}. They are grown on a Si(100)/SiO$_x$ (100 nm) high resistive substrates with a Ta 3-nm seed layer. The stack of interest is Pt(5)/Co($t_{Co}$)/Au(2.5)/Co($t_{Co}$)/Pt(5) (all thicknesses are given in nanometer)\cite{SI}. In such a heterostructure, magnetic textures (domain walls or skyrmions depending on the magnetic parameters) with opposite chirality in each layer are coupled by dipolar interaction and are sensitive to spin-orbit torques induced by the charge to spin conversion in both Pt outer layers. The Pt/Co interface provides a large DMI\cite{Yang_PRL} with $D_s = 0.66\pm0.1$~pJ/m\cite{Hrabec_PRL,Gehanne_PRR}. The Pt layer thickness enables an efficient charge to spin conversion and therefore a large spin-orbit torque gets induced to the Co layers. The Au thickness is sufficient to avoid any exchange coupling between the Co layers\cite{Hrabec_PRL,V-Grolier_PRL} that are only coupled through dipolar interaction. In this symmetric stack, the chirality is not only given by the DMI but also by the flux closure at the domain walls or transition regions\cite{Bellec_EPL,Hrabec_Nat.Comm.}, which increase the total chiral energy\cite{Hrabec_PRL}. The Co thicknesses are tuned between 1.30 and 1.60 nm in order to modify the magnetic anisotropy. As measured on samples with a single Co layer, the interface ($K_S/t_{Co}$) and shape ($-\frac12\mu_0M_S^2$) magnetic anisotropies compensate each other at $\sim$1.65--1.70~nm, where the effective magnetic anisotropy\cite{winter1961} $K_\mathrm{eff}=K_S/t_{Co}-\frac12\mu_0M_S^2$ is close to zero. Using $M_S=1.55\times10^6$~A/m$^2$, the interfacial anisotropy energy is therefore estimated to $K_S = 2.55\pm0.10$~mJ/m$^2$.\cite{SI}
The Co thickness is varied by steps of 0.5~{\AA}. While such a step is below the atomic layer thickness, the magnetic layer roughness is not impacted, since the layers are not epitaxial and consist of grains of 10--15~nm diameter. Similarly, the texture pinning landscape (relevant  for the skyrmion motion, as well as for the texture morphology evolution) is not affected by the roughness eventually induced by partial atomic layers, since it is mostly related to the granular structure, as demonstrated in earlier studies\cite{gross2018}.
Current induced nucleation and motion was studied in 1 to 3 $\mu$m tracks obtained using e-beam lithography and Ar$^+$ etching, connected by Ti/Au contacts. Magnetic textures were observed using magnetic force microscopy using home-built soft MFM tips with a CoCr/Cr bilayer coating to reduce tip induced perturbation of magnetic textures~\cite{Chauleau_PRB,torrejon2012,Hrabec_Nat.Comm.,SI}.

\section{Skyrmion stabilization at zero applied field}

Magnetic textures arise from the equilibrium between all the micromagnetic energies. To describe skyrmions, the two key parameters are the domain wall energy $\sigma=4\sqrt{AK_\mathrm{eff}}-\pi D$\cite{Rohart_PRB} ($A$ is the exchange energy constant) and the long-range dipolar coupling\cite{Kooy_PRP,Boulle_Nat.Nano.,Buttner_Sci.Rep.,B-Mantel_PRB}. Large DMI lower the domain wall energy and can destabilize the uniform state for $D>4\sqrt{AK_\mathrm{eff}}/\pi$.\cite{Rohart_PRB} Long-range dipolar couplings promote a demagnetized state by creating a flux-closure situation between up and down magnetization regions and provide another mechanism to create magnetic textures. However, such an effect is limited in ultrathin films, and beyond a film thickness $\sigma/\mu_0M_S^2$ their influence is not significant\cite{Kooy_PRP,Boulle_Nat.Nano.}. These two criterions provide a guideline for material optimization through the magnetic film thickness, so that the relative energy of uniform and non-collinear magnetization phases can be tuned.

\begin{figure}[ht]
\includegraphics[width=\columnwidth]{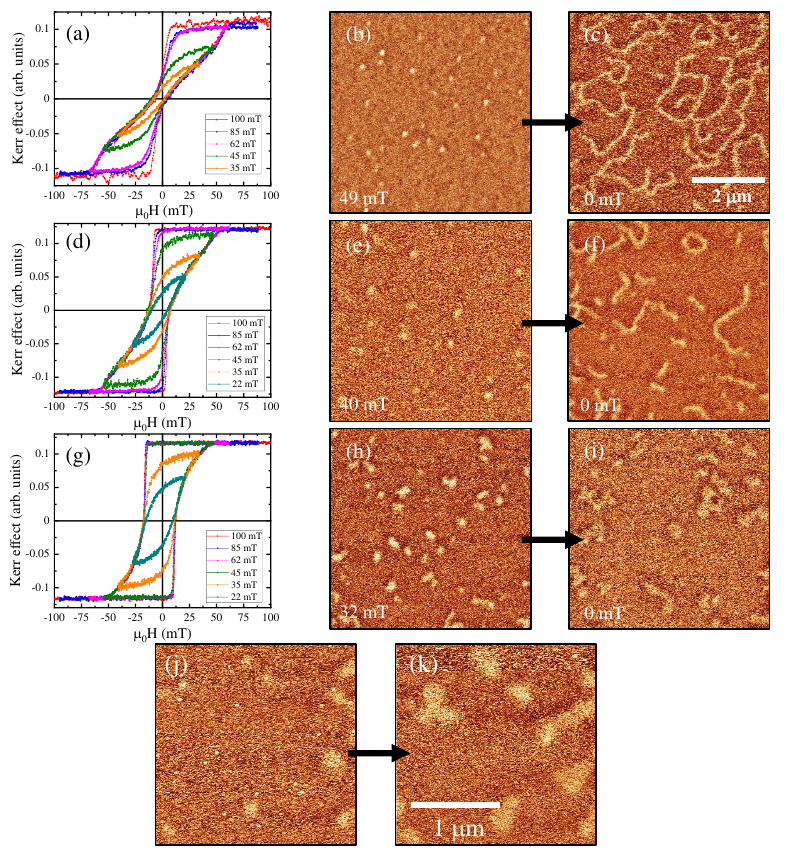}
    \caption{
    Hysteresis loops (a,d,g), field stabilized skyrmion states (b,e,h,j) and following relaxed state in zero field (c,f,i,k) MFM images, for different Co thicknesses [(a-c) $t_{Co}=1.6$~nm, (d-f) $t_{Co}=1.55$~nm, (g-i) $t_{Co}=1.5$~nm]. The hysteresis loops are taken for different maximum fields as shown in the legends, to display full and minor loops. The field stabilized skyrmion images are taken close to the magnetic saturation, respectively at $\sim$49 (b), 40 (e) and 32~mT (h). They display skyrmions with $\sim$85-150~nm (b),110-170 nm (e) and 150-300 nm (h) diameters respectively. All images have the same size, indicated in (c). Images (j) and (k) correspond to a closer view of the $t_{Co}=1.5$~nm sample, both taken exactly at the sample place, and show how the textures relax at zero magnetic field, from a field stabilized state.
    }
    \label{Fig1}
\end{figure}

Close to magnetic anisotropy compensation thickness, where $K_\mathrm{eff}\approx0$, the samples spontaneously demagnetize, as shown in Fig.~\ref{Fig1}(a), for a $t_{Co}=1.60$~nm thick sample, since both criterions defined above are satisfied.\cite{Boulle_Nat.Nano.,Hrabec_Nat.Comm.} In such samples, the remanent state always corresponds to a disordered stripe phase (see image in supplemental material\cite{SI}), whatever the magnetic history, as underlined by the minor loops in Fig.~\ref{Fig1}(a). This result is a direct consequence of the low effective magnetic anisotropy and therefore low domain wall energy. Under the action of the DMI and/or the dipolar couplings, the creation of stable domains with a reversed magnetization is energetically favorable. Textures can proliferate, hence the destabilization of the uniform magnetization state. Upon the application of a perpendicular magnetic field close to the saturation, isolated 85 to 150 nm diameter skyrmions are observed (Fig.\ref{Fig1}(b)), with a core oriented antiparallel to the applied field. However, when releasing the external field, these skyrmions turn out to be unstable. As shown in Fig.~\ref{Fig1}(c), they spontaneously expend to form worm-like domains, in order to maximize the length of domain walls, an effect similar to  the run-out instability introduced to described magnetic bubbles~\cite{Thiele_JAP}. Note that this remanent state is slightly different from the true demagnetized state, which is attributed to a small energy barrier to nucleate additional stripes.

Decreasing the Co thickness increases the effective magnetic anisotropy. From 1.55~nm and below, the hysteresis loop have a $\sim$100\% remanent magnetization, showing that the stability of the uniform state is improved. Below 1.20 nm, square loops are observed, which indicates that the magnetic anisotropy is so large that up or down uniform states are the only stable states~\cite{Hrabec_PRL}. The transition from zero remanent to square loops is however continuous, with a rich variety of loops observed between 1.30 and 1.55~nm\cite{SI}. In this thickness range, despite the 100~\% remanence, the magnetic state does not switch abruptly to the reversed uniform state, but to intermediate states, that contain magnetic textures, as shown in Fig.~\ref{Fig1}(e) and (h) for $t_{Co}=1.55$ and 1.50~nm. This is interpreted as a  thickness range with an intermediate domain wall energy, sufficiently large to avoid self-demagnetization at zero field but that still enables magnetic textures. This is also underlined by the fact that these samples can easily be demagnetized through conventional demagnetization protocol and minor loops as shown in Fig.~\ref{Fig1}(d) and (g).
Close to the magnetic saturation, isolated skyrmions are stabilized for cobalt thicknesses of 1.50 and 1.55~nm. In these states, the skyrmion core is antiparallel to the applied magnetic field, which maintains a small and compact shape.

From the field stabilized skyrmion state, releasing the applied magnetic field shows different behavior. In the 1.55~nm thick sample, the domain wall energy is  still too low to prevent a run-out instability, which results in a zero field magnetic state similar to that in thicker samples, although with a more limited domain expansion.
On the contrary for $t_{Co}=1.5$~nm (see in Fig.~\ref{Fig1}(g-i)), the larger anisotropy and domain wall energy prevent the instability and  skyrmions remain rather compact in zero field. The main evolution is a diameter increase, since the applied field was compressing to skyrmion core. However, small deformations (moderate anisotropic expansion, flower-like shapes...) cannot be avoided, which shows that the texture morphology is close to be unstable. Even if they probably keep to skyrmion topology, like any domains in chiral magnetic samples, they are certainly too fragile to be used as zero-field skyrmions. Note however that the long term stability of these states at zero field is good, as checked by imaging the same sample after several hours without any noticeable change\cite{SI}.

It is tempting to further decrease the Co thicknesses to improve the shape stability at zero magnetic field. Indeed, the larger magnetic anisotropy should induce a larger domain wall energy and should suppress the  run-out instability, following the trend observed from 1.60 to 1.50 nm Co thickness. Between 1.30 and 1.45~nm Co thicknesses, the hysteresis loops are not perfectly square and show that some intermediate textured states are possible. However, only stripes or worm-like states could be observed during the field process, with an abrupt transition from stripes to the homogeneous state when increasing the field (see loops and images in the supplemental material\cite{SI}). In these samples, the stabilization of skyrmions using a field sweeping procedure looks impossible.

\begin{figure}[ht]
\centering
\includegraphics[width=1.0\linewidth]{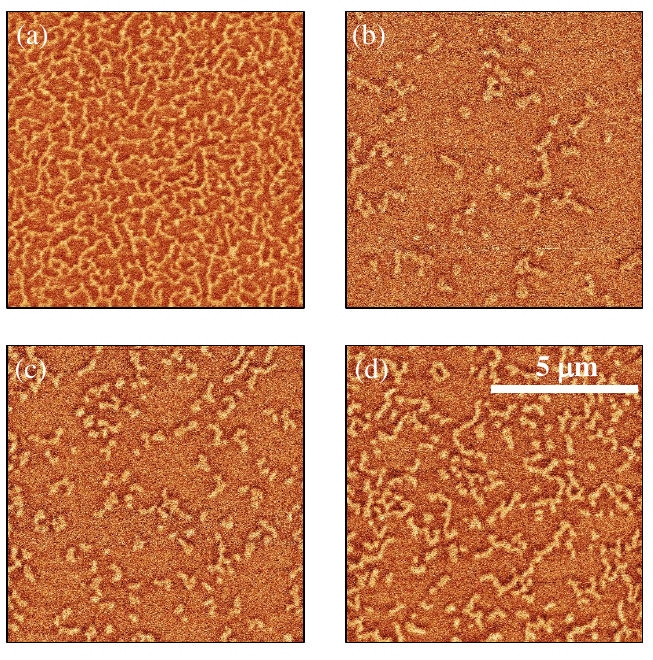}
\caption{(a)-(d) MFM images captured at magnetic remanence in the bilayer sample with $t_{Co}= 1.50$ nm. Image (a) is obtained after a standard demagnetization procedure. Images (b-d) are obtained at zero field, following a saturation in negative field and the application of 32, 28, and 24 mT respectively.}
\label{Fig2}
\end{figure}

Samples with a 100~\% remanence enable the stabilization of  different magnetic states  at zero field (density and morphology changes), depending on the magnetic history, as visible in the minor loops in Fig.~\ref{Fig1}d and g. Contrary to a sample where the homogeneous state is not stable in zero field, here, the textures stabilized at a given applied field remain relatively stable at zero field, without any parasitic nucleation of new textures. This is illustrated in the $t_{Co}=1.50$-nm sample, in Fig.~\ref{Fig2}. Using a standard demagnetization process (sinusoidal field sweeping, with a progressive amplitude reduction), a disordered stripe phase, similar to that observed in self-demagnetizing sample, is obtained [Fig.~\ref{Fig2}(a)]. After the application of a field smaller than the saturation field, different states are stabilized at remanence. They correspond to a continuous evolution (isotropic and anisotropy expansion) of the textures stabilized at the applied field. Note that, the larger is the last applied field the more compact are the textures at remanence.

\section{Zero field skyrmion current control}

In the previous section, skyrmion-like magnetic textures have been stabilized at zero applied magnetic field from an existing skyrmion state close to the saturation field. Provided that no run-out instability occurs at zero field, this process is straightforward involving a continuous evolution from the field stabilized texture, without any modification of the topology. On the contrary, skyrmion nucleation directly from the uniformly magnetized state is a more challenging process. Indeed, a topology change is required and no specific final state is promoted.
When a field is applied close to the magnetic saturation, textures that contain magnetic moments oriented antiparallel to the applied field are so unfavored that skyrmions are generally the most likely excited state, as demonstrated in several experiments\cite{Fallon_Small,Jiang_Science,Hrabec_Nat.Comm.}. At zero applied magnetic field, the variety of possible magnetic states, as underlined by the observations in Fig.~\ref{Fig2}, leads to more complexity and uncertainty. It is therefore highly demanding on the material optimization, so that skyrmions become the most favorable nucleated result. Beyond an obvious applied interest, such a nucleation experiment provides a test of the different accessible states and their relative energy.

This problematic is investigated using current-induced nucleation in the vicinity of sharp electrodes in magnetic tracks\cite{SI}, using a protocol previously developed for nucleation at finite fields~\cite{Hrabec_Nat.Comm.}. We consider devices with 1 or 3 $\mu$m wide tracks, connected by Ti/Au contacts. On one side the contact consists of a sharp electrode, whereas on the other one the contact is flat across the wire and do not provide any current concentration. At the sharp contact current concentration, current line divergence, heating, and spin accumulation may occur and participate to the nucleation\cite{Heinonen_PRB, Gorchon_PRL,Lemesh-Adv.Mat.,Litzius_Nat.Elec.}. However, considering the symmetry of the multilayer stack and current injection geometry, the Oersted field associated to the current is negligible close to the contact, where skyrmions are created~\cite{Hrabec_Nat.Comm.} and should play a limited role.
Before each set of measurements, the device is saturated using a perpendicular magnetic field and short 2 to 7-ns-long current pulses are applied in the absence of any external field.

Even if zero field skyrmion-like texture stabilization is possible for $t_{Co}=1.50$~nm from a field procedure, the application of a current pulse leads to a demagnetization of the whole track and the formation of a stripe phase. Indeed, in these samples, the nucleation energy barrier is so small that the global track heating easily destabilizes the uniform state, a result similar to the work of Lemesh et al.~\cite{Lemesh-Adv.Mat.}. Improved results could be obtained with thinner samples. For samples with $t_{Co}=1.35$~nm, the application of current pulses ($J=0.4\times10^{12}$~A/m$^2$, 5 ns) leads to the formation of stripes only in the vicinity of the sharp contact. This indicates that the nucleation energy barrier is increased as compared to the previous samples and that textures can be nucleated only in the hottest part of the sample. In this sample, skyrmions are not obtained, since the stripe phase is likely to have a lower energy, but this result still indicates an increase in texture energy. This is coherent with the approach developed in previous part where the domain wall energy is increased by decreasing the Co thickness. The application of current pulses to the  $t_{Co}=1.30$~nm sample leads to the nucleation of compact skyrmionic textures in the vicinity of the sharp contact. In this sample, the larger domain wall energy prohibits the elongation of the texture after the formation of a small nucleus and the compact shape is retained. Note that this result is obtained in a sample where the thickness is so small that the hysteresis loop is almost square,\cite{SI} indicating that any texture has an energy significantly larger than the homogeneous state. The difference between the $t_{Co}=1.30$~nm sample and the thicker samples underlines that, here, the skyrmion state is well separated in energy from the other texture, making their nucleation the most likely result, independently on the precise nucleation mechanism.

\begin{figure}[ht]
\centering
\includegraphics[width=1.0\linewidth]{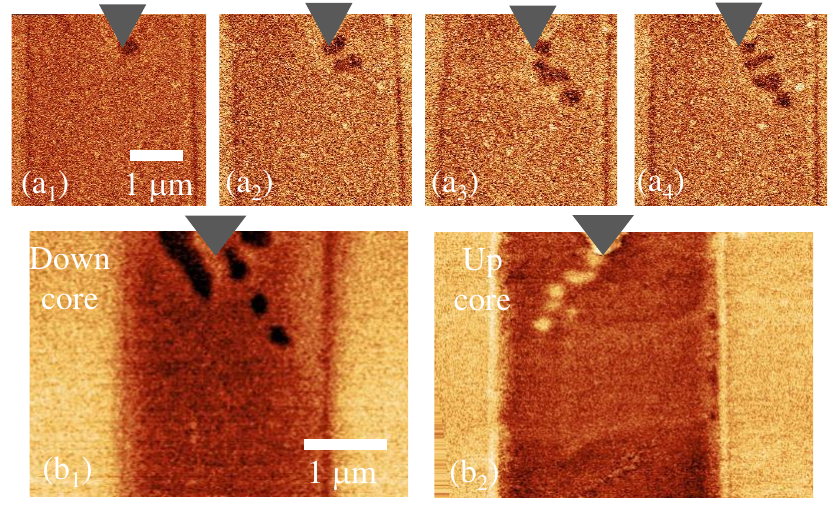}
\caption{(a1) - (a4) Current induced nucleation in the vicinity of a sharp contact (triangular notch in top of each image) on the $t_{Co} = 1.30$-nm bilayer samples, in a 3 $\mu$m wide track. Starting from a saturated state (not shown), a single current pulse ($J = 0.85\times10^{12}$~A/m$^2$, 4 ns) is applied between two successive images. (b1) and (b2) show the nucleation of skyrmions followed by application of 3 (b1: down core) and 4 (b2: up core) current pulses ($J = 0.85 \times 10^{12}$~A/m$^2$, 5 ns), respectively. }
\label{Fig3}
\end{figure}

Fig. \ref{Fig3}(a1 - a4) show the skyrmion nucleation experiments in the $t_{Co}=1.30$~nm sample, using a 3~$\mu$m wide track. Single current pulse of $J=0.85\times10^{12}$~A/m$^2$, 4 ns is applied between two successive images, starting from a uniform state (all spins saturated towards up orientation: not shown here). The current flows from the sharp electrode (represented by the grey triangle on the top of each image) towards the flat electrode. Compact magnetic textures appear in the vicinity of the electrode with a core polarity (dark), opposite to the initial saturation direction, as expected for any excited texture over a uniform magnetic state. The dark contrast corresponds to an attractive force, hence a skyrmion core parallel to the microscope tip magnetization. Fig.~\ref{Fig3}(a1 - a4) show a succession of images to evidence that the process is progressive. In this experiment, each pulse nucleates exactly one skyrmion and pushes the existing ones forming a train of almost equally spaced skyrmions. Further, (b1) and (b2) show the nucleation of skyrmions followed by application of 3 current pulses ($J=0.85\times10^{12}$~A/m$^2$, 5 ns), but starting from initial uniform states with opposite magnetic orientation. The skyrmions that are nucleated hence have opposite core polarization, as indicated by the opposite contrast in the MFM images. This demonstrates that the process of nucleation is independent on the core polarization. The nucleated skyrmions are deflected toward the right or the left respectively, depending on the skyrmion core polarization. This behavior is a direct consequence of the topological character of the skyrmions. Due to their particular topology, skyrmion traveling at a velocity $\mathbf{v}$ undergo a gyrotropic force $\mathbf{G}\times\mathbf{v}$, with $\mathbf{G}=(M_S t_{Co}/\gamma)S\mathbf{z}$ ($\gamma$ is the gyromagnetic ratio), that induces a deflection, whose direction depends on the topological number $S$. Since opposite skyrmion core leads to opposite topological number \cite{Tokura_Review}, this force explains the opposite deflection direction.

\begin{figure}
\centering
\includegraphics[width=1.0\linewidth]{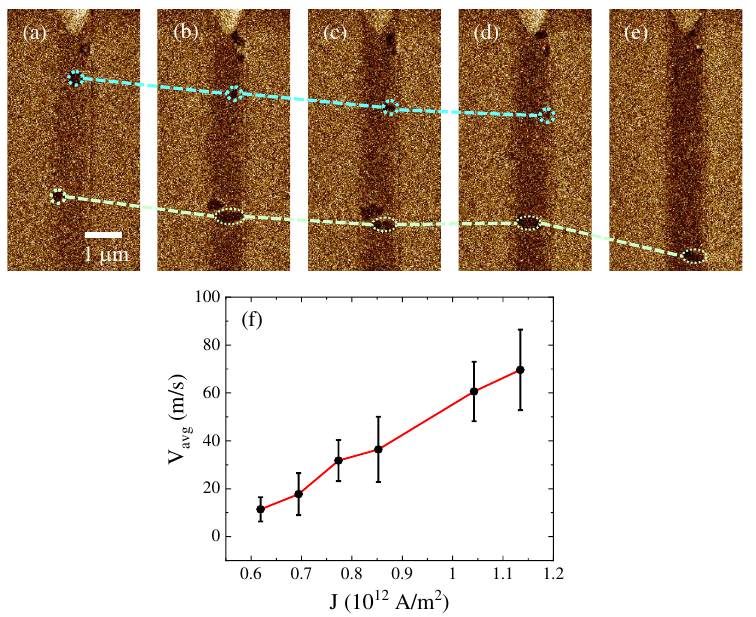}
\caption{(a)-(e) show the motion of skyrmions under the influence of a series of current pulses ($J=1.13\times10^{12} A/m^2$, 4 ns) for the sample with $t_{Co}= 1.30$ nm. The dashed circles and lines represent the positions and movements of the skyrmions, respectively. (f) shows the velocity ($v_{avg}$) vs current density ($J$) plot for sample with $t_{Co}= 1.30$ nm. The error bar in (f) corresponds to the difference of velocities between the fastest and slowest skyrmions in between two successive electrical pulses.}
\label{Fig4}
\end{figure}

The skyrmion current induced motion in the $t_{Co}=1.30$-nm-thick sample has been further studied in 1-$\mu$m-wide tracks where the skyrmions are better confined. Current pulses with varying intensity $J$ and duration from 4 to 10 ns were used to move  the skyrmions at zero applied field. Fig.\ref{Fig4}(a)-(e) show the motion of skyrmions under the influence of a series of current pulses ($J=1.13\times10^{12} A/m^2$, 4 ns). The dashed circles and lines represents the positions and movements of the skyrmions, respectively. The direction of current pulse is from the point contact towards to the flat electrode (top to down) in the sample. The skyrmions are pushed along the current direction, which suggests that the motion is induced by the spin-Hall effect in the Pt layers, as expected considering that the Pt/Co interface DMI induces a left-handed chirality ~\cite{Hrabec_Nat.Comm.,Yang_PRL, Pizzini_PRL,Belmeguenai_PRB}. It can be observed that the movement of the skyrmions in between two pulses is not uniform and eventually differs from one skyrmion to the other. This is associated to the role of defects and skyrmion motion mostly occurs by hopping within the inhomogeneous pinning landscape of the sample, which limits their velocity \cite{Kim_Nature}.
Further, the skyrmions either annihilate or remain pinned once they reach the side of the magnetic tracks. Fig.\ref{Fig4}(f) shows the average velocity $v_{avg}$ vs current density. A pinning threshold of about $0.5\times 10^{12}$~A/m$^2$ has been observed, a bit higher than the best published results, probably due to the high pinning at the Co/Au interface\cite{Jeudy_PRL}. However, the skyrmion mobility is comparable to the best results with a maximum average velocity of $\sim$70 m/s, obtained at $1.13\times10^{12}$~A/m$^2$.  Application of current pulses with larger $J$ leads to skyrmion deformations or further nucleation within the track. At the largest current densities, a stripe phase is formed in the whole track. Skyrmion motion under few other current densities and pulses are presented in supplemental material\cite{SI}.

\section{Conclusion}

In this work, we have shown that, magnetic skyrmions can be stable in the absence of an external magnetic field, if the sample micromagnetic parameters, are finely tuned via the sample thickness.
Contrary to usual approaches that enable field stabilized skyrmions, this requires samples with a finite perpendicular magnetic anisotropy to prevent the skyrmions from being converted to elongated structures (run-out instability). The ability to stabilize skyrmion-like textures from field process is found to be an insufficient condition to enable their use in skyrmion-based spintronic devices. Indeed, when excited by current, some skyrmions can be deformed to elongated structures. Robust skyrmions whose shape remain compact during the motion  require an even larger anisotropy, and therefore an accurate optimization process. In these conditions, the skyrmions behave as solitonic excitations. We have demonstrated the ability to nucleate and move with electrical current pulses zero field skyrmions, with an efficiency similar to the best established results of the literature using field stabilized skyrmions. Beyond an obvious applied interest for applications, this demonstration opens a new possibility to stabilize skyrmions and paves a road toward their control in antiferromagnetic films, where a magnetic field cannot be applied to the samples.

\acknowledgments{
We acknowledge fruitful discussions with Andr\'e Thiaville.
This work was supported as a Indo-French collaborative project supported by CEFIPRA (IFC/5808-1 / 2017), by the French National Research Agency (ANR), under Contract No. ANR-17-CE24-0025 (TopSky)  and a public grant overseen by the ANR as part of the ``Investissements d'Avenir'' program (Labex NanoSaclay, reference: ANR-10-LABX-0035, SPICY).
}


\end{document}